\documentstyle[12pt,fps]{article}
\makeatletter
\@addtoreset{equation}{section}
\makeatother

\newcommand{\bibi}{\bibitem}

\newcommand{\eq}{\ref}
\newcommand{\beq}{\begin{equation}}
\newcommand{\eeq}{\end{equation}}
\newcommand{\bea}{\begin{eqnarray}}
\newcommand{\eea}{\end{eqnarray}}

\newcommand{\cc}{\cite}
\newcommand{\lb}{\label}

\newcommand{\chb}{\overline{\chi}}

\newcommand{\half}{\frac{1}{2}}

\newcommand{\al}{\alpha}
\newcommand{\bt}{\beta}
\newcommand{\lag}{\langle}
\newcommand{\rag}{\rangle}
\newcommand{\gm}{\gamma}

\newcommand{\kp}{\kappa}
\newcommand{\lm}{\lambda}

\newcommand{\sg}{\sigma}

\newcommand{\vr}{\varphi}

\newcommand{\ps}{\psi}

\newcommand{\psb}{\overline{\ps}}

\newcommand{\dmu}{\partial_{\mu}}
\newcommand{\dMu}{\partial^{\mu}}

\newcommand{\dsl}{\partial \!\!\!/}

\newcommand{\ageq}{\makebox[0mm][l]{\raisebox{-0.45ex}{\small$\sim$}}\mbox{\raisebox{+0.5ex}{\small$>$}}}
\newcommand{\ra}{\rightarrow}

\newcommand{\be}{\begin{equation}}
\newcommand{\ee}{\end{equation}}

\def \3{\ss}

\def\dateandnumber(#1)#2#3#4{
\vbox to 18mm{%
     \hbox to \textwidth{ \hspace*{14mm} \hsize=40mm%
            \vbox{%
                 \hbox to 40mm{\large #1 \hss}%
                 \hbox to 40mm{    \hss}%
                 \hbox to 40mm{    \hss}%
                 }%
                 \hss \hsize=80mm%
            \vbox{%
                 \hbox to 80mm{\hss \large #2}
                 \hbox to 80mm{\hss \large #3}
                 \hbox to 80mm{\hss \large #4}
                 }%
            \hspace*{14mm} }%
      \vss
    }
}
\def\titleofpreprint#1#2#3#4{{\LARGE \bf
\vbox to 43mm{%
     \vss
     \hbox to \textwidth{ \hspace*{14mm} \hsize=130mm%
            \hss \vbox{
                      \hbox to 130mm{\hss \LARGE \bf #1\hss}%
                      \hbox to 130mm{\hss \LARGE \bf #2\hss}%
                      \hbox to 130mm{\hss \LARGE \bf #3\hss}%
                      \hbox to 130mm{\hss \LARGE \bf #4\hss}%
                 }%
            \hss \hspace*{14mm} }%
      \vss
    }
}}

\def\listofauthors#1#2#3{{\large
\vbox to 22mm{%
     \vss
     \hbox to \textwidth{ \hspace*{14mm} \hsize=130mm%
            \hss \vbox{
                      \hbox to 130mm{\hss \large #1\hss}%
                      \hbox to 130mm{\hss \large #2\hss}%
                      \hbox to 130mm{\hss \large #3\hss}%
                 }%
            \hss \hspace*{14mm} }%
      \vss
    }}
}
\def\listofaddresses#1#2#3#4{{\small
   \vbox to 18mm{%
        \vss
        \hbox to \textwidth{ \hspace*{14mm} \hsize=130mm%
               \hss \vbox{
                         \hbox to 130mm{\hss \small #1\hss}%
                         \hbox to 130mm{\hss \small #2\hss}%
                         \hbox to 130mm{\hss \small #3\hss}%
                         \hbox to 130mm{\hss \small #4\hss}%
                         }%
               \hss \hspace*{14mm}
        }%
        \vss
   }}
}
\def\abstractofpreprint#1{{\normalsize
\vbox to 110mm{%
     \vss
     \hbox to \textwidth{\hss \normalsize \bf Abstract \hss}%
     \normalsize
     #1
     \vss
     }}
}
\def\footnoteoftitle#1{{\small
\vbox to 30mm{\parindent0pt
     \vss\small #1 \vss
    }}
}
\def\footnoteitem(#1)#2{
\begin{list}{#1}{\labelwidth4.0mm \leftmargin7.0mm
\labelsep2.5mm \rightmargin7.0mm \parsep0.5ex plus0.2ex minus0.1ex
\itemsep0ex plus0.2ex }
\item #2
\end{list}
}
\begin{document}
\dateandnumber(   )%
{J\"ulich, HLRZ 111/93}%
{                    }%
{                    }%
\titleofpreprint%
{                On the equivalence between                    }%
{             2D Yukawa and Gross--Neveu models                }%
{                                                              }%
{                                                              }%
{                                                              }%
\listofauthors%
{          E.~Focht$^{1,2}$, W.~Franzki$^{1,2}$,               }%
{       J.~Jers\'ak$^{1,2}$ and M.A.~Stephanov$^3$             }%
{                                                              }%
\listofaddresses%
{\em \mbox{}$^1$Institute of Theoretical Physics E,             %
     RWTH Aachen,        D-52074 Aachen, Germany               }%
{\em \mbox{}$^2$HLRZ c/o KFA J\"ulich,                          %
     , D-52425 J\"ulich, Germany                               }%
{\em \mbox{}$^3$ Theoretical Physics, 1 Keble Rd.,              %
            Oxford OX1 3NP, UK                                 }%
{                                                               %
                                                               }%
\abstractofpreprint{
We study numerically on the lattice the 2D Yukawa model with the U(1)
chiral  
symmetry and $N_F$ = 16 at infinite scalar field self-coupling. 
The scaling behaviour of the fermion mass, as the Yukawa coupling
approaches zero, is analysed using the mean field method. It is found  
to agree with that of the Gross-Neveu model with the same symmetry and 
$N_F$. This is so even if the sign of the bare
kinetic term  of the scalar field is 
negative. This suggests that the 2D Yukawa models belong to the 
universality class of the Gross-Neveu models not only at weak scalar 
field self-coupling but also for a broad range of the bare parameters 
which is not accessible to the $1/N_F$ expansion. 
New universality classes might arise at the crossover to the spin 
model universality class, however. 
}
\footnoteoftitle{
\footnoteitem($^*$){ \sloppy
Supported
by Deutsches Bundesministerium f\"ur Forschung und Technologie,
by Deutsche Forschungsgemeinschaft and by Jesus College, Oxford.
}
}
\pagebreak

%

\section{Introduction}

The 2D Gross-Neveu  (GN$_2$)  models with various global chiral
symmetries \cc{GrNe74} have been of
continuous interest since it has been realized that these models are
asymptotically free and that the dynamical
mass generation (DMG) occurs \cc{GrNe74,Wi78b}.
The 2D Yukawa (Y$_2$) models with the same symmetries have received very little
attention, however, though they can potentially have very similar properties.
These models are defined in the continuous Euclidean space by the action
\be
   S _{cont} = \int d^2x \left[ \sum_\mu \dmu\vr^* \dMu\vr + \half m_0^2|\vr|^2
   + \frac{\widetilde{\lm}}{4!} |\vr|^4 + \psb\dsl\ps +
   \widetilde{y} \psb(\mbox{Re}\vr+i\gm_P\mbox{Im}\vr)\ps \right]
   \lb{CONTACTION}
\ee
where $m_0$, $\widetilde{\lm}$ and $\widetilde{y}$ are the bare scalar mass
and  bare dimensionful coupling constants, respectively. For
definiteness we have written down the model with chiral U(1) symmetry
and a quartic scalar field selfcoupling. The generalization to other
symmetries and/or other selfcouplings is obvious. The fermion field
$\psi$ consists of $N_F$ flavour components and $\vr$ is a complex
scalar field.

 It was noticed long ago \cc{LuMa64} that the Yukawa models in various
dimensions reduce to the four-fermion models if the bare
kinetic and selfinteraction terms for the scalar field vanish. This
relationship in 4d was used in the models of composite Higgs boson 
\cc{BaHi91}. Such a relationship is
a rather simple kinematical fact which is even more clearly seen in lattice
regularized Yukawa theories \cc{Shi89a,HaHa91}.

 In this paper we present numerical results suggesting in 2D a much
more profound relationship,
namely that the Y$_2$ field theory belongs 
in a broad range of parameters to one universality class, which is the class
 of the GN$_2$ model with the same symmetry.
 Such  equivalence might be indicated by the dimensionality of the
coupling parameters in the
Y$_2$ models, but we are aware of no clear demonstration on such a
general level. 
 For large $N_F$ and small $\widetilde{y}$, $\widetilde{\lm}=O(1/N_F)$
the equivalence can be shown in the large $N_F$ expansion \cc{Zi91}.
Already fifteen years ago the equivalence between  Y$_2$ and GN$_2$
has been suggested by Guralnik and Tamvakis
\cc{TaGu78} on the basis of a mean field (MF) analysis for a wide
class of scalar selfcouplings.
However, the  MF method is traditionally considered not to be reliable in
2D. 

 Recently we have found numerical evidence that in analogy to the
GN$_2$ models 
the Y$_2$ models with the Z(2) and U(1)  symmetry
on the lattice are asymptotically free in the Yukawa coupling
even if the selfcoupling of the scalar field is arbitrarily strong
\cc{DeFo93a,DeFo93b}.
 This property is suggested by the MF method \cc{DeFo93b} and in the
effective potential approach
\cc{She93} which is in spirit similar to the MF approximation.
It is argued in ref. \cc{DeFo93b} that the MF method is applicable to
the Y$_2$ models because the fermions induce long range effective
interaction which couples many scalar variables ferromagnetically.
Such interaction can be well described by a mean field.  The local
selfinteractions of the scalar field (without fermions) are treated
exactly or numerically. 
Such a version of the MF method is applicable on finite
lattices and gives a good agreement with the numerical data. This
allowed us to demonstrate the exponential decrease of the fermion mass
with $1/\widetilde{y}^2$, and thus the asymptotic freedom of the Y$_2$
models.

 The agreement between the  numerical and MF methods obtained until
now suggests to test
numerically for various values of the coupling parameters and
with improved data
the MF inspired conjecture \cc{TaGu78,She93,DeFo93b} that
the Y$_2$ and the GN$_2$
models with the same symmetry belong to the same universality class,
i.e. correspond to the
same renormalized quantum field theory and have the same physical
content in a broad range of the coupling parameters.

 For this purpose we have investigated quantitatively the rate with
which the fermion 
mass $am_F$ in lattice units ($a$ being the lattice constant)
approaches zero as the bare Yukawa coupling $\widetilde{y}$ decreases.
We compare this $\widetilde{y}$-dependence with the MF predictions for
various fixed values of 
the dimensionless bare
parameters $a^2m_0^2$ and $a^2\widetilde{\lm}$ .
 The parameters  should be fixed such that the pure scalar model at
$y=0$ is in the high temperature phase (symmetric phase in the Z(2)
case, vortex phase in the U(1) case, etc.).
 The expected asymptotic scaling law in the infinite volume is
\beq
am_F \sim \exp \left( -\frac{1}{2\bt_0}\frac{1}{g^2}\right)~,
\lb{Y2:SCALING}
\eeq
where
\beq
g^2 = \chi y^2,
\eeq
$y$ is the bare dimensionless Yukawa coupling of the
lattice Y$_2$ model and
$\chi$ is the susceptibility of the pure scalar model at $y=0$.
For positive values of the hopping parameter $\kp$ of the scalar field
(its definition is given below) one has
\bea
  \chi & = & \frac{Z_\phi}{(am_\phi)^2}, \lb{CHIEQ} \\
  y^2 & = & 2\kp a^2\widetilde{y}^2,
\eea
$am_\phi$ being  the renormalized mass of the scalar boson in the high
temperature phase and $Z_\phi$ its wave function renormalization constant.

 We find numerically that in the U(1) model with $N_F=16$ the
coefficient $\bt_0$ is within certain accuracy independent both of
$am_\phi$  
and of the strength of the bare selfcoupling of the scalar field.
 This is so even if this coupling approaches infinity.
 The values of $\bt_0$ obtained from the numerical simulation are
close to the value of the  first
coefficient of the $\bt$-function of the corresponding GN$_2$ model.
 The agreement is good (within 7\%)  at large $am_\phi$ and is
found to be even better
if the sign of the bare kinetic term of the scalar field is chosen negative.
 On the other hand, for small $am_\phi$ significant deviations from the MF prediction 
are observed in the range of couplings and lattice sizes we have investigated.
 This might indicate a crossover towards the scaling properties of the
pure scalar theory.

The parameter region corresponding to the negative sign of the bare
scalar kinetic term (negative $\kp$) is of particular interest.
 Here, far outside of the scope of the perturbation expansion,
the contributions from the fermion determinant to the effective kinetic term
of the scalar field make the Y$_2$ models apparently field
theoretically as consistent as the GN$_2$ models.

 Further support for the conjecture of the universality of the Y$_2$
and GN$_2$ theories is provided
by our numerical results for the ratio of the mass of the scalar
fermion-antifermion bound state $B$ and of the fermion 
mass, $m_B/m_F$.
The mass $am_B$ is extracted from the two-point function of the
composite operator $\psb\ps$.
 The ratio is found to be independent of the chosen values of the
bare parameters of the Y$_2$ model. 
For  $N_F=16$ fermions it is close to 2 and in agreement with the
predicted \cc{DaHa75,She76} value 1.99 in the GN$_2$ model.

 The outline of the paper is as follows: In the next section we define
the lattice Y$_2$ models and describe the structure of their phase
diagram, as implied by the MF method and our previous numerical results.
The various universality classes, suggested by this structure, are
discussed.
 In sec. 3 we show how the universality of the Y$_2$ models can be
studied by means of the scaling behavior of the fermion mass. Here we
also describe how this scaling behavior can be determined, using the MF
method, on lattices of modest sizes and at rather large values of the
Yukawa coupling.
 Section 4 presents numerical evidence obtained within the U(1)
symmetric Y$_2$ model with $N_F=16$
that the scaling behavior of the
fermion mass is very close to that of the GN$_2$ models. Some
differences are noted too, however, and interpreted as a sign of a
cross-over to the spin-model universality classes.
 In sec. 5 we further support the idea of equivalence between the Y$_2$
and GN$_2$ models by comparing the ratio of the two lowest masses.
 Section 6 and the appendix contains some analytic speculations about
the cross-over between the GN$_2$ and spin-model universality classes.
A brief summary is presented in sec.~7.

\section{Phase diagram of the Y$_2$ model}

We investigate Yukawa models with Z(2) and U(1) chiral symmetry using 
mostly staggered fermions. The lattice actions have been chosen
such that the GN$_2$ and the spin models are their simple special cases.
E.g. the action of the U(1) model for staggered fermions is:
\begin{equation}
S=S_B+S_F+S_Y
\lb{ACTION}
\end{equation}
\begin{equation}
S_B=\sum_x \sum_i \left[-2\kappa\sum_\mu\phi_x^i\phi_{x+\mu}^i
   +\phi_x^i\phi_x^i+\lambda(\phi_x^2-1)^2 \right]
\end{equation}
\begin{equation}
S_F=\frac{1}{2}\sum_{x,\alpha,\mu} \left( \overline{\chi}_x^\alpha
    \eta_{\mu x} \chi_{x+\mu}^\alpha - \overline{\chi}_{x+\mu}^\alpha
    \eta_{\mu x} \chi_x^\alpha \right)
\end{equation}
\begin{equation}
S_Y=y\sum_{x,\alpha}\overline{\chi}_x^\alpha \sum_{b\in P}\left(
\phi_{x-b}^1+i(-1)^{x_1+x_2}\phi_{x-b}^2 \right) \chi_x^\alpha
\lb{ACTIONYUK}
\end{equation}
where $\phi$ is the two-component scalar field and $\chi^\alpha$,
$\alpha=1,\dots,N$, are $N$ staggered fermion fields, which describe
in the continuum limit $N_F=2N$ Dirac fermions. We chose the
representation  $\eta_{1x}=(-1)^{x_2}$ and $\eta_{2x}=1$ for the sign
factors. The dimensionless bare parameters are
$\kappa$, the scalar hopping parameter, $\lambda$, the scalar
quartic selfcoupling and $y$, the hypercubic Yukawa coupling, which
couples the scalar fields on a plaquette $P$ to the fermionic fields
located in one corner of $P$. The action for the Z(2) model is similar,
$\phi$ are then only 1--component scalar fields.

The continuum action (\eq{CONTACTION}) arises from the lattice action
(\eq{ACTION}) if $a\ra 0$ and the following rescalings are made:
\beq
  \renewcommand{\arraystretch}{1.5}
  \begin{array}{rcl}
     \vr             & = & \sqrt{2\kp}\phi \\
     m_0^2           & = & (1-2\lm-4\kp)/(a^2\kp) \\
     \widetilde{\lm} & = & 6\lm/(a^2\kp^2) \\
     \widetilde{y}   & = & y/(a\sqrt{2\kp}) 
  \end{array}
  \lb{CLTRAFO} 
\eeq
The components of $\psi$ arise from those of $\chi/\sqrt{a}$ in a
complicated but standard way.

In the case $\kappa=\lambda=0$ this
action describes the Gross--Neveu models on the lattice
with $\phi$ being an ``auxiliary''
scalar field. The usual GN$_2$ coupling constant is 
\be
g=y/\sqrt{2}.
\lb{GN:COUPLING}
\ee
 The parametrization used in (\eq{ACTION}) makes it obvious
that the GN$_2$ models are special cases of the Y$_2$ models.
In the parametrisation usual in the continuum (\eq{CONTACTION}) the
case $\kp=\lm=0$ corresponds to the nonperturbative limit $a^2m_0^2$,
$a^2\widetilde{y}^2\ra\infty$, $\widetilde{y}^2/m_0^2=g^2$ fixed.

 The other well understood limit is $y=0$,
where the action (\eq{ACTION}) describes the pure 2--component scalar
field theory and free massless fermions. In the limit
$\lm=\infty$ it is the 2D XY model. In the Z(2) case the 2D Ising
model is obtained in this limit.

The $1/N_F$ expansion usually used for the GN$_2$ model
can be
applied to the Y$_2$ models only for $\lambda=O(1/N_F)$ and leads to
results which are  equivalent, at least in the leading $1/N_F$ order,
 to the GN$_2$ case \cc{Zi91}.
In \cc{DeFo93b} we argued that the MF approximation can be used 
to describe the scaling behavior and also some finite cutoff
and finite size effects in the Y$_2$ models for any $\lm$.
 The fermion determinant generates nonlocal ferromagnetic
interactions between
$\phi$'s.
The MF approximation is applicable because a large number of sites
contribute to the mean field.  

 The method is briefly  as follows.
The effect of the fermion determinant can be described by a mean
field $H(\sg,y)$ acting on $\phi$ as an external field,
\beq
H(\sg,y)=N y^2 \sg \int \frac{d^2p}{(2\pi)^2} \left( \sum_\mu \sin^2 p_\mu +
(y\sg)^2 \right)^{-1},
\lb{MEANFIELD}
\eeq
where $\sg$ is the mean magnetization. 
 The mean field (\ref{MEANFIELD}) acts as an external field on the scalar
model given by the action $S_B$ from (\eq{ACTION}) and the
magnetization $\langle\phi\rangle$ is given by the selfconsistency
equation
\beq
 \lag\phi\rag = f(H(\lag\phi\rag,y)),
 \lb{SELFCONSIST}
\eeq
and $am_F=y\lag\phi\rag$. Here $f(H)$ is the response function of the
scalar  model.
On a finite
lattice the momentum integrals in $H(\lag\phi\rag,y)$ are replaced by
sums over lattice momenta. The resulting equation is solved numerically
by recursion predicting $\lag\phi\rag$ as a function of the
Yukawa coupling $y$, the scalar couplings $\kp$, $\lm$ and the volume
of the system $L^2$.
 In order to obtain the response function $f(H)$
we study the scalar model, with interactions from the pure scalar part of
the action, in the external field $H$.
This we can do numerically and in some special cases exactly. The mean field $H$ can then be determined selfconsistently.

We note that the logarithmically diverging nonlocality of the
ferromagnetic interaction producing the mean field in (\eq{MEANFIELD})
makes the DMG possible at arbitrarily weak Yukawa coupling and leads
to the asymptotic freedom.

As discussed in \cc{DeFo93b} the MF approximation leads also to a better
understanding of the fermion mass generation mechanism in the case of
the Y$_2$ model with continuous chiral symmetry which cannot be
broken spontaneously according to the Mermin--Wagner--Coleman theorem. 
The main idea is that the effective interaction induced by the fermions
can produce ferromagnetic ordering of scalar variables on distances
$O(1/m_F)$. On larger distances the ordering is destroyed by the long
wavelength fluctuations (spin waves) as usual in two dimensions.
The scalar field $\phi$ in a volume of a linear size
$O(1/m_F)$ has a nonzero magnetization $\langle\phi\rangle_{1/m_F}$
which direction is drifting slowly. The mass of the fermion coupled to
such a scalar field is given by $am_F \simeq
y\langle\phi\rangle_{1/m_F}$. We checked this relation and found a good
agreement in our Monte Carlo simulations. One can view this relation
as a selfconsistency equation for $m_F$. Alternatively 
it can be used as a formal definition of $\langle\phi\rangle_{1/m_F}$.
Thus we can apply the MF approach to the
Z(2) and U(1) models in complete analogy if we bear in
mind that by $\sg$ we mean $\langle\phi\rangle_{1/m_F}$.

%
%
\begin{figure}
\centerline{
 \epsfxsize=7cm
 \epsfbox{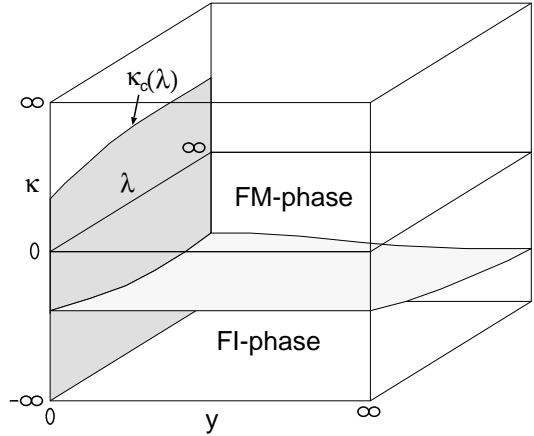}
}
\caption{Sketch of the phase structure of 2 dimensional Yukawa models
with hypercubic Yukawa coupling in the 3 dimensional coupling
parameter space.}
\label{PHASEDIAGRAM}
\end{figure}

Figure \ref{PHASEDIAGRAM} is a schematic representation of the phase
diagram of the Y$_2$ models with Z(2) or U(1) chiral symmetry. At
$y=0$ we have a critical line
$\kappa_c(\lambda)$ separating two phases of the pure scalar field
theory. In the Z(2) case these are the high
temperature, paramagnetic (PM) phase and the low temperature,
ferromagnetic (FM) phase. In the U(1) case the corresponding phases
are called
vortex (VX) phase and spin wave (SW) phase, respectively, alluding to
the dominant spin configurations.

 There is considerable evidence provided by some exact, semiclassical
and large $N_F$ results \cc{GrNe74,DaHa75,She76,Wi78b,FoNi91} that
the GN$_2$ models are asymptotically free and exhibit the DMG. 
This implies that on the line $\kappa=\lambda=0$ the systems
are for arbitrarily small coupling $y$  in
the broken (FM) phase for the Z(2) symmetry or in the SW phase in the
U(1) case. The
models have an essential singularity at $y=0$. Analytic and numerical
investigations of the Yukawa models \cc{TaGu78,DeFo93a,She93,DeFo93b}
indicate that for $\kappa<\kappa_c(\lambda)$ the theories behave
for any $\lambda$ similarly to the GN$_2$ models. They are asymptotically
free. At arbitrarily small $y$ the models are in the FM or
SW phase for the Z(2) or U(1) case, respectively. The surface
$\kappa<\kappa_c(\lambda)$, $y=0$ is a critical surface at which the
masses of the GN$_2$ model vanish, if expressed in lattice units. 
One can perform 
the continuum limit approaching any point of this surface. 
For some lattice discretizations of the
GN$_2$ or Yukawa models on  the PM or VX phases can appear
at large $y$ \cc{Af82,DeFo93a}. This is a lattice artifact, which can
be avoided by using the hypercubic Yukawa coupling.

The scalar part of the action (\ref{ACTION}) has the so-called staggered
symmetry
\be
\phi_x \rightarrow (-1)^{x_1+x_2}\phi_x =: \phi_x^{st},~
\kappa\rightarrow -\kappa.
\ee
In the scalar models this symmetry implies
a phase transition line at $\kp=-\kp_c(\lm)$ which separates
the PM and the antiferromagnetic phase in the Z(2) model and the
VX phase from the staggered spin wave phase in the U(1) case.

The staggered scalar field $\phi^{st}$ does not couple to the fermions
through the
hypercubic Yukawa coupling (\eq{ACTIONYUK}). Therefore we expect the
phase transition at
$\kp=-\kp_c(\lm)$ to occur also for $y>0$ and
to depend only weakly on $y$. It forms a
critical sheet which separates in the Z(2) case the FM from the FI
(ferrimagnetic) phase (see fig. \ref{PHASEDIAGRAM}).
We have checked the existence of the FI phase in the case of the Z(2)
model numerically. A similar critical sheet is expected also in the
U(1) model.

Though we
do not know whether the critical sheet $y=0$, $\kp \le \kp_c(\lm)$ extends to
$\kappa=-\infty$ or not, we see no reason why it shouldn't. 
 The phase transition around $\kp\simeq-\kp_c(\lm)$ does not influence
$am_F$ which seems to scale with $y\ra 0$ also below this sheet.

The described phase structure suggests the existence of several
universality classes in the Y$_2$ models. They depend on
the position of the critical point where one performs the continuum limit
and, as we shall see,
in some cases probably also on the way how this point is approached.

When one performs the continuum limit on the critical line
$\kappa_c(\lambda)$ ($\lambda>0$) and approaches it within the plane
$y=0$, in the Z(2) case the theory belongs to the Ising universality class. 
In the U(1) case the transition is Kosterlitz--Thouless-like, the
theory belongs to the universality class of the 2D XY model.
At $\lambda=0$ and $\kappa=1/4$ there is the Gaussian fixed point. When
approaching this point from smaller $\kappa$ at $y=0$ we have a theory
with free noninteracting scalars and free massless fermions.
The continuum limit towards the line $\kp=\kp_c(\lm)$ taken from the
$y>0$ region is rather complex and will be discussed in section 6.

Approaching the critical surface $y=0$, $\kappa<\kappa_c(\lambda)$,
leads to an asymptotically free theory
\cc{TaGu78,DeFo93a,She93,DeFo93b}. But it is not a priori clear whether
this theory is for any $\lm>0$ and $\kp<\kp_c(\lm)$
in the same universality class as the GN$_2$ model at $\lm=\kp=0$.
At least the
spectrum is expected to be similar, with fermions and fermionic bound states
scaling while the masses of the states in the $\phi^4$ theories, in
particular $am_\phi$, remain nonzero
and decouple in such a continuum limit. The discussion of this limit is the
main aim of the next two sections of this paper. We shall concentrate
on two aspects of universality: the question whether the theories
scale in the same way when one approaches the critical surface at different
($\kappa,\lambda$) points. More precisely, whether the first coefficient
of the $\beta$ function, which can be interpreted as a critical
exponent, is the same. We also examine the ratio of the first two
masses in the spectrum of the Y$_2$ model and compare it with that of
the GN$_2$ model.

We note that the structure of the phase diagram and the continuation of
the critical surface at $y=0$ to negative $\kp$ suggests to consider also
$\kp<0$, though the rescalings (\eq{CLTRAFO}) are not well defined. Even
if negative $\kp$ corresponds to the bare scalar kinetic term with wrong
sign, the continuum Y$_2$ theories can be physically sensible at $\kp<0$.
A strong indication for such a phenomenon has been found numerically in
the 4D Yukawa theories on the lattice for sufficiently strong Yukawa
coupling \cc{Y4}. Thus we shall study  the scaling
behavior at negative $\kp$ also.

At the critical surface around $\kp\simeq -\kp_c(\lm)$, $y>0$, the
fermion mass does not scale and fermions would decouple in the
corresponding continuum limit.

\section{Scaling behaviour of $am_F$ and its analysis on finite lattices}

The asymptotic scaling law for the GN model (1 loop) is
\beq
am_F \sim \exp \left( -\frac{1}{2\bt_0}\frac{1}{g^2} \right).
\lb{GN:SCALING}
\eeq
We note that the correct values of $\bt_0$ obtained in perturbation
theory are
\beq
\bt_0=\frac{N_F-1}{2\pi} \qquad \mbox{(Z(2))},
\qquad \bt_0=\frac{N_F}{2\pi} \qquad \mbox{(U(1))}.
\lb{BETAS}
\eeq

The MF considerations \cc{DeFo93b,St94} and the study of the
effective potential \cc{She93} predict that the fermion masses in
the more general case of the Y$_2$ models behave according to the
scaling law (\eq{Y2:SCALING}).
Here $\chi=\chi(\kp,\lm)$ is the susceptibility of the pure scalar
model described by the part $S_B$ of the action (\ref{ACTION}).
For $0<\kp<\kp_c(\lm)$ it is given by (\eq{CHIEQ}), but the scaling law
(\eq{Y2:SCALING})  holds also for $\kp\leq 0$. 
Indeed, $\chi$ is finite for such $\kp$ values.
In particular,
\be
\chi(\kp,0)=\frac{1-4\kp}{2\kp}.
\ee
The predicted value $\bt_0=N_F/2\pi$ does not depend on $\kp$  and $\lm$ and
thus coincides with the value of the first coefficient of the $\bt$
function of the GN$_2$ model in eq. (\eq{GN:SCALING}), as predicted by
the MF approximation in that case.
In the U(1) case the MF approximation gives $\bt_0$ correctly. We note
that the scaling law (\eq{GN:SCALING}) of the GN$_2$ model agrees
with (\eq{Y2:SCALING}) because of  eq. (\eq{GN:COUPLING}) and
$\chi(0,0)=1/2$. 

\begin{figure}
\centerline{
 \epsfxsize=5cm
 \epsfbox{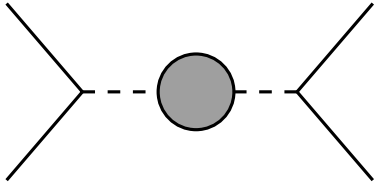}
}
\caption{The 4 point function in a Yukawa theory in the
lowest order in $y$ and the scalar selfinteraction fully included.}
\label{VERTEX}
\end{figure}

 As mentioned in the previous section we concentrate on the question
whether the continuum limit taken at the critical surface $y=0$,
$\kappa<\kappa_c(\lambda)$ is everywhere Gross--Neveu like. Therefore we
want to compare the scaling behaviour of the fermion mass $am_F$ for
$y\ra 0$ at various fixed $\kappa$ and $\lambda$.
 In order to make such a comparison it is necessary first to compare 
the strength of the Yukawa coupling itself.
The interacting scalar field namely modifies the effective strength
of the Yukawa interaction already on the fermionic tree level.
This is obvious when the fermion four-point function $G^{(4)}_F$ is
considered
in the lowest order in y, namely O($y^2$), but with the scalar
selfinteraction fully included, as shown in Fig. \ref{VERTEX}.
At zero momentum we obtain 
\be
     G^{(4)}_F(p_i=0) = y^2 G_\phi^{(2)}(p=0) = y^2 \chi.
\ee
Thus we see that the scaling behaviour should be compared
with respect to the effective strength of the Yukawa coupling
\be
    g^2 = \chi y^2.
\ee
(We thank F. Niedermayer for an elucidating discussion on this
point.)
Therefore the scaling behavior of $am_F$ for various fixed $\kp$, $\lm$ values
is interpreted as the same if the numerically determined coefficient
$\bt_0$ in the scaling law (\eq{Y2:SCALING}) is the same. Its
consistency with the values (\eq{BETAS}) is then an indication that the
Y$_2$ models belong for those $\kp$, $\lm$ values to the GN$_2$
universality class.

%
%

The aim of our numerical investigations is thus to check whether the
scaling law (\ref{Y2:SCALING}) is consistent with the data.
However, eq. (\ref{Y2:SCALING}) is valid only in the infinite
volume limit at very small $y$. 
In ref. \cc{DeFo93b} we described how the MF method can make
predictions about the fermion masses on finite lattices. 

\begin{figure}
\centerline{
 \def\fpsangle{90}
 \epsfxsize=10cm
 \fpsbox{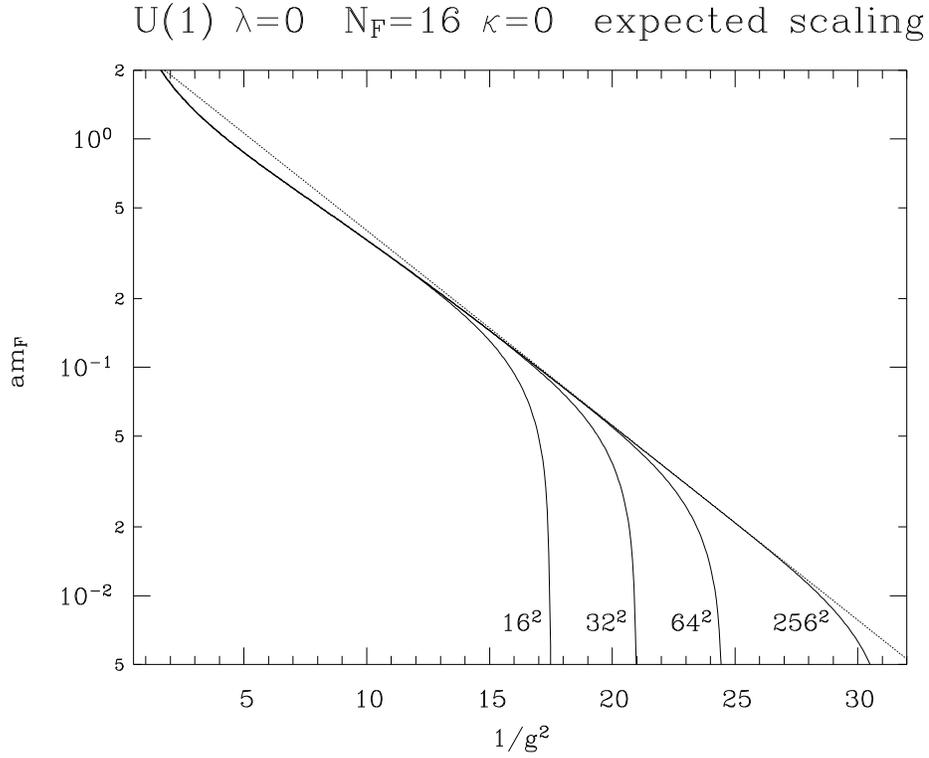}
}
\caption{Mean field approximation prediction for the fermion mass in
the GN case, plotted logarithmically against $1/g^2=1/(\chi y^2)$. The
slope of the straight line corresponds to the asymptotic scaling, eq.
(\protect\ref{GN:SCALING}).} 
\label{GN:MFP}
\end{figure}

These MF predictions allow us to estimate how Yukawa models approach
the asymptotic scaling. In fig. \ref{GN:MFP} we have plotted the fermion
mass $am_F$ obtained by solving eq. (\ref{SELFCONSIST}) for $\kp=\lm=0$
on lattices with $L=16,~32,~64$ and $256$. The onset of asymptotic
scaling represented by the straight line
is around $1/g^2\approx 20$, i.e. it can be seen only on very
large lattices ($L \ageq 200$).

%
%

%
%
\begin{figure}
\centerline{
 \def\fpsangle{90}
 \epsfxsize=10cm
 \fpsbox{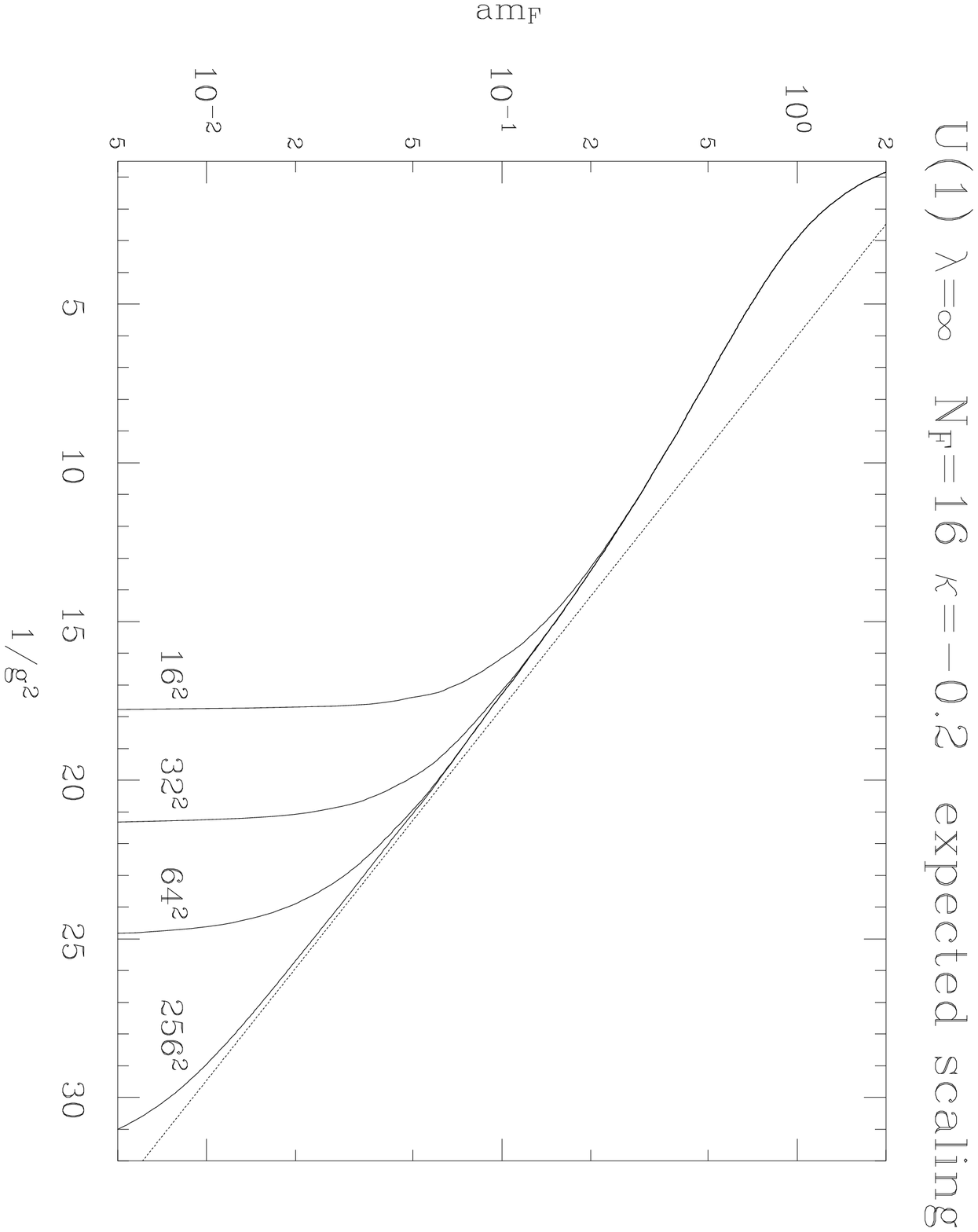}
}
\caption{Same as fig. \protect\ref{GN:MFP}, now for $\lm=\infty$,
$\kp=-0.2$.}
\label{LINF:MFP}
\end{figure}

Figure \ref{LINF:MFP} shows the MF prediction for $am_F$ at
$\lm=\infty$ and $\kp=-0.2$. The picture is similar to the GN$_2$
case, but an indication of the same
asymptotic scaling seems to require even larger lattices.
For other $\kp$ at $\lm=\infty$ the situation is analogous.

The MF analysis thus suggests such a slow approach to the asymptotic
scaling that we have no chance to achieve it in numerical simulations.
Too small $y$ and, correspondingly, too large lattices would be needed.
The strategy we adopt in this situation is the following: we compare the
numerical data for $am_F$ with the MF predictions for finite volumes and
correspondingly large $y$. If we find an agreement within a reasonable
accuracy margin, we conclude that the MF method works well
and that its
asymptotic predictions are thus correct.
In other words, the MF analysis of the data is
the theoretical means  
for extrapolation to large $1/g^2$ and $L$ of the data
obtained numerically at moderate values of these parameters.

To compare numerical data with the MF predictions we introduced
an ad-hoc correction $r$ to be determined in fits to the data. 
It's role is to take into
account eventual deviations from the mean field approximation and to
quantify the (dis)agreement between data and the prediction.
It is introduced in the following way: instead of $am_F=y\lag\phi\rag$
we use in eqs. (\ref{MEANFIELD}), (\ref{SELFCONSIST}) $am_F=r y
\lag\phi\rag$ and fit for the parameter $r$. 
In this MF motivated Ansatz the asymptotic scaling would have a
different form, namely
$$
am_F\sim e^{-\frac{1}{2\bt_0r}\frac{1}{g^2}}.
$$
However, on lattices much smaller  than the sizes at which the
asymptotic behavior sets on, $r$ parametrizes the difference between
the MF prediction of the approach to the asymptotic scaling and the
true one. So we expect $r\ra 1$ as $L\ra\infty$ and $y\ra 0$.
 The result $r=1$ would
mean perfect agreement between the data and the MF prediction for the
coefficient $\bt_0$ in the asymptotic scaling behavior  in eq.
(\eq{Y2:SCALING}), 
which of course cannot be expected on finite lattices.

\section{Observed $y$ dependence of $am_F$}

In order to verify the hypothesis that GN and Y$_2$ models are in the
same universality class as $y\ra 0$
for $\kp<\kp_c(\lm)$ we have studied the $y$-dependence of $am_F$ in the
U(1) model with $N_F=16$. The numerical data
have been collected using staggered fermions
on lattices of the sizes
$L=16$, $32$, $64$ at many $y$ points
for five pairs of fixed $\kp$ and $\lm$ values (see table 1).
 Using the hybrid Monte Carlo algorithm, the fermion propagators have
been measured in coordinate and in momentum 
space and have been fitted with the free fermion Ansatz. The results
for $am_F$
are very stable and the errors so small that 
the $y$-dependence of this mass clearly shows deviations from the
MF predictions on finite lattices.

%
%
\begin{figure}
\centerline{
 \def\fpsangle{90}
 \epsfxsize=10cm
 \fpsbox{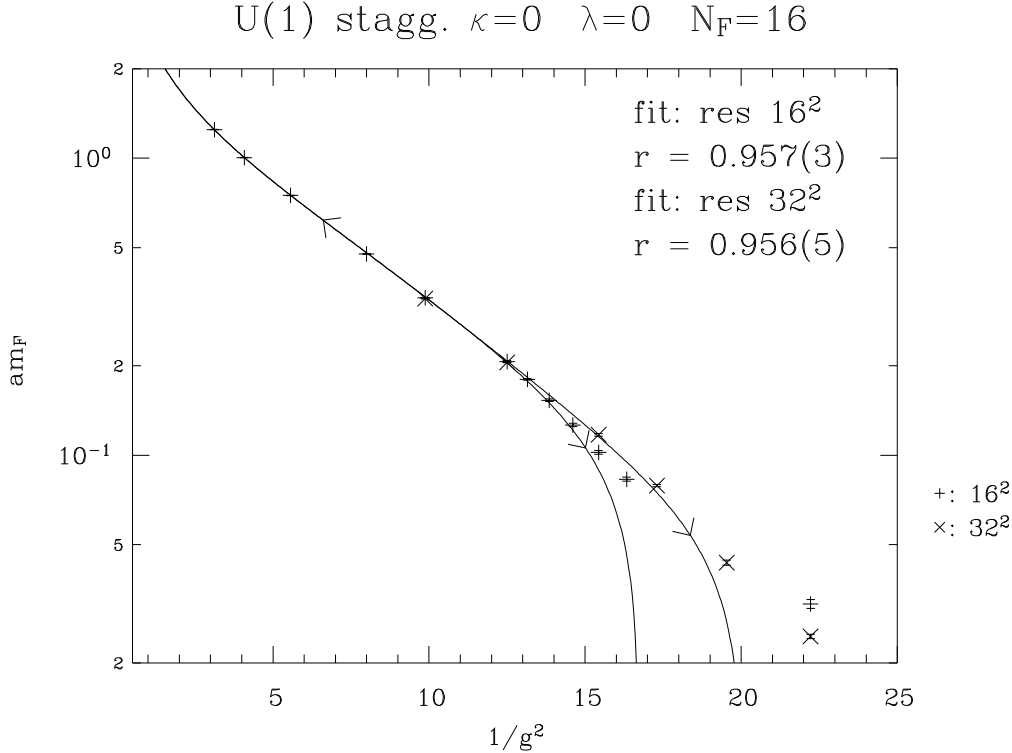}
}
\caption{Numerical data for fermion mass $m_F$ on finite lattices  in
the U(1) Y$_2$ model plotted against $1/g^2$, where $g$ is the effective
Yukawa coupling $g=\protect\sqrt{\chi}y$.}
\label{GN:FIT}
\end{figure}

%
%
\begin{figure}
\centerline{
 \def\fpsangle{90}
 \epsfxsize=10cm
 \fpsbox{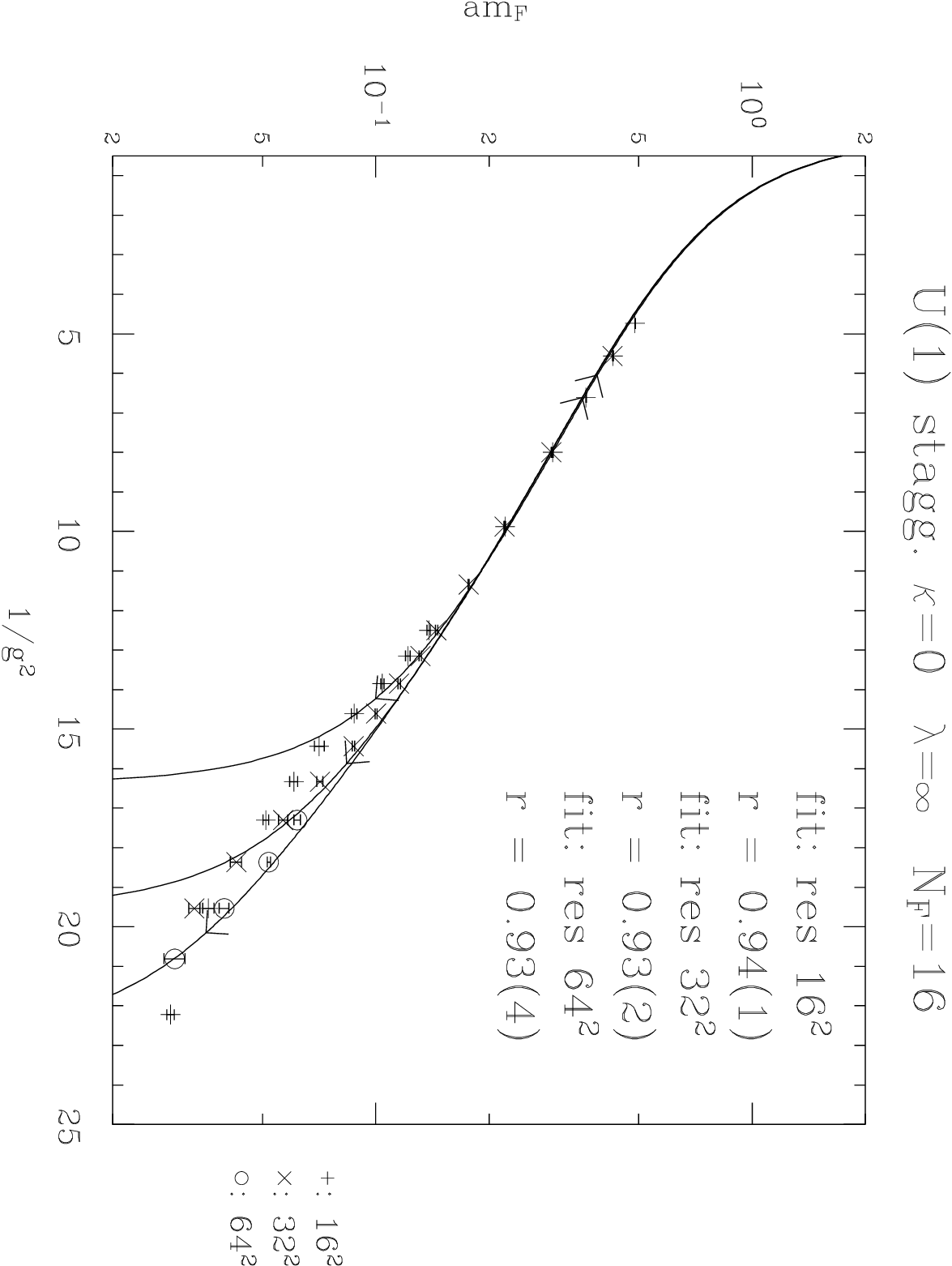}
}
\caption{Same as fig. \protect\ref{GN:FIT}, now for $\lm=\infty$ and
$\kp=0$.}
\label{Y1:FIT}
\end{figure}

%
%
\begin{figure}
\centerline{
 \def\fpsangle{90}
 \epsfxsize=10cm
 \fpsbox{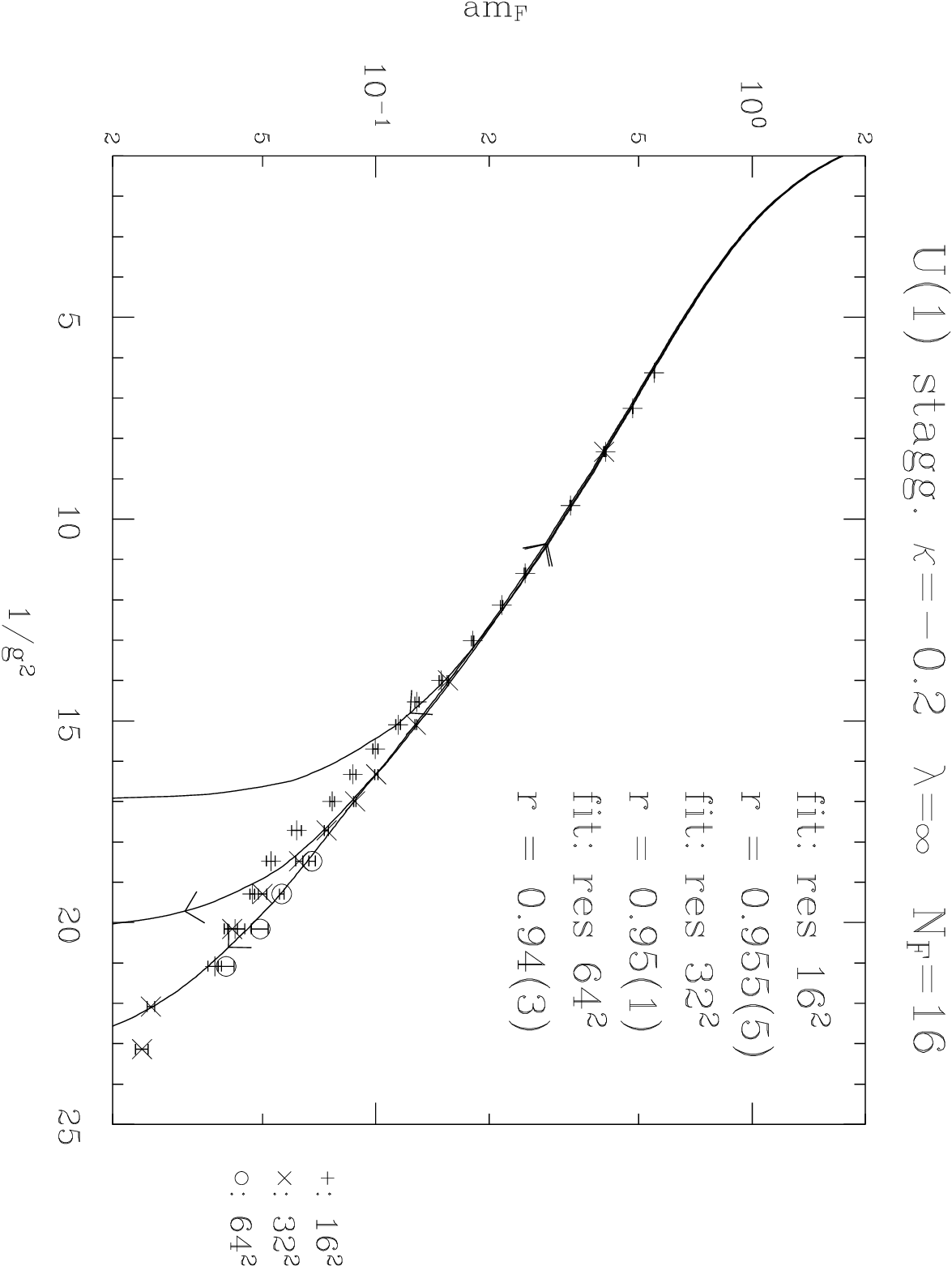}
}
\caption{Same as fig. \protect\ref{GN:FIT}, for $\lm=\infty$ and
$\kp=-0.2$.}
\label{Y2:FIT}
\end{figure}

%
%
\begin{figure}
\centerline{
 \def\fpsangle{90}
 \epsfxsize=10cm
 \fpsbox{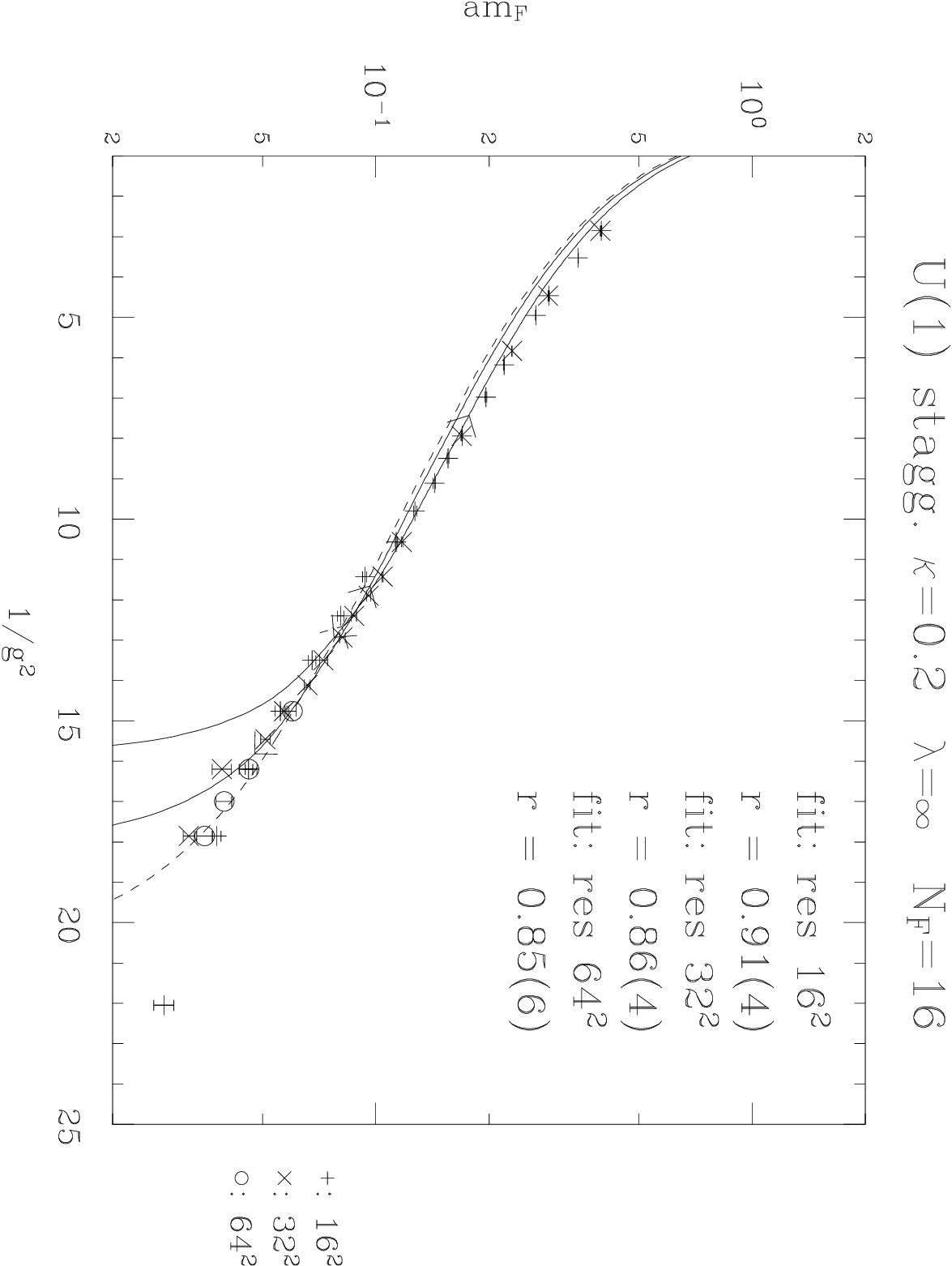}
}
\caption{Same as fig. \protect\ref{GN:FIT}, for $\lm=\infty$ and
$\kp=0.2$.}
\label{Y3:FIT}
\end{figure}

The pair $\kp=\lm=0$ (the GN$_2$ model, which is  the best understood
case) has been chosen to estimate
the quality of the MF description of the data.
 For this purpose we have determined the value of the parameter $r$ by
a fit to the data as described at the end of the preceding section.
 The fit is shown in fig. 5 and we find that $r$ has the
same values for both lattice sizes $L=16,32$ we have used in this case
and deviates from $1$ by $4.5\%$.
Having chosen rather large $N_F=16$ we can be sure that the GN$_2$
model scales according to eq. (\eq{Y2:SCALING}), as predicted by the
large $N_F$ expansion.
So the deviations of $r$ from 1 of this magnitude should not be
interpreted as a signal 
for a scaling behavior different from (\eq{Y2:SCALING}), but rather as
an estimate of the accuracy within which the MF method works on finite
lattices.

\begin{table}
 \begin{center}
  \begin{tabular}{|cc|ccc|}
   \hline
   $\kp$ & $\lm$ & $16^2$ & $32^2$ & $64^2$ \\
   \hline
   0   & 0        & 0.957(3) & 0.956(5) & \\
  -0.4 & $\infty$ & 0.971(5) & 0.970(7) & \\
  -0.2 & $\infty$ & 0.955(5) & 0.95(1)  & 0.94(3) \\
   0   & $\infty$ & 0.94(1)  & 0.93(2)  & 0.93(4) \\
   0.2 & $\infty$ & 0.91(4)  & 0.86(4)  & 0.85(6) \\
   \hline
  \end{tabular}
 \end{center}
 \caption{Values of the fit parameter $r$ in the U(1) Y$_2$ model with
   $N_F=16$ on $16^2$, $32^2$ and $64^2$ lattices.}
 \label{R:TAB}
\end{table}

In the figs. \ref{Y1:FIT}, \ref{Y2:FIT} and \ref{Y3:FIT} we present
numerical results of the hybrid MC simulation of the Y$_2$ model
for $\lm=\infty$ and $\kp=0$, $-0.2$ and $+0.2$, respectively. The
results for the fit 
parameter $r$ are shown in the table \ref{R:TAB}.
We have concentrated on $\lm=\infty$ because this value is the opposite
extreme to the case $\lm=O(1/N_F)$ when (\eq{Y2:SCALING}) follows from
the $1/N_F$ expansion. Thus an agreement with
(\eq{Y2:SCALING}) at $\lm=\infty$ allows to expect that a similar
agreement can be found for any intermediate value of $\lm$.

The agreement is indeed found for $\kp\leq 0$.
At $\kp=0$ (fig. \ref{Y1:FIT}) we have $r\approx 0.93$. The fits are
good and although the deviation of $r$ from $1$ is larger than in the GN
case (fig. \ref{GN:FIT}), it is still similar to that case. The fit in
fig. \ref{Y2:FIT} is done for data at $\kp=-0.2$ and the results for $r$
are the same as in the GN model (within errors). For $\kp=-0.4$ the
value of $r$ is even nearer to 1 than in the GN case (tab. \ref{R:TAB},
figure not included).
There is a clear sign, that $\bt_0$ is the same for $\kp=\lm=0$ and
$\kp\leq 0$, $\lm=\infty$.

 At $\kp=0.2$, $\lm=\infty$ the fit in fig. \ref{Y3:FIT} shows
considerable deviation of $r$ from 1, however. The quality of the fit
is also poor, i.e. the data cannot be described by
the MF Ansatz in spite of the presence of the $L$-dependent parameter
$r$. 
 We suppose that this is an effect of the crossover region to
the other universality classes at $\kp=\kp_c(\lm)$ (discussed in
section 6).

The presented results for the $y$-dependence of $am_F$ 
are only a part of the data for $am_F$ 
we have accumulated during our studies of the Y$_2$ models.
We have investigated also the U(1) symmetric lattice Y$_2$ model with
naive fermions \cc{DeFo93a,DeFo93b} and for $\lm=0$ and $\lm=0.5$
performed also simulations of the Z(2) model with various $N_F$.
Though less detailed, these results are consistent with those
presented here. Put together, our results strongly suggest that, as
$y\ra0$ at fixed $\lm\geq 0$ and  $\kp$ satisfying
$-\kp_c(\lm)<\kp<\kp_c(\lm)$, the fermion mass $am_F$ behaves
according to the scaling law (\eq{Y2:SCALING}). We interpret this as a
piece of evidence that the Y$_2$ models belong in the indicated region
of $\lm$, $\kp$ parameters to the same universality class as the
GN$_2$ models with the same $N_F$ and chiral symmetry.

We note that neither the present data nor the MF method of their
analysis allow to determine the values of the mass gap to compare with
the exact result \cc{FoNi91}.
For this purpose
either the data in the asymptotic scaling region or a much better
analysis of finite cut-off effects would be needed.

\section{The mass of the $\overline{\chi}\chi$ state}

Needless to say, more observables should be compared if the
equivalence of various field theoretical models is under discussion.
 Besides critical exponents, the simplest universal quantities are the
ratios of masses. A comparison of the spectrum can give valuable
hints about the universality classes of similar theories.

Early analytic investigations \cc{DaHa75,She76} have revealed a rich
spectrum in the GN$_2$ model. The mass of the first excited state, which is interpreted as a fermion--antifermion bound state, is
\beq
      m_B = 2m_F \cos\left(\frac{\pi}{2(N_F-1)}\right),
\lb{MB}
\eeq
followed by other fermion--antifermion bound states
and multifermion bound states.

We could determine the mass $am_B$ in the lattice units. 
We followed the methods of ref. \cc{AtBe90} and
measured the time-dependent correlation function
\beq
C_B(x_2) = \left\lag \frac{1}{L} \sum_{x_1} | M_{0x}^{-1} |^2
\right\rag~,
\lb{PROP:DEF}
\eeq
where $x_1$ is the Euclidean time, $x_2$ the space coordinate and
$M_{0x}^{-1}$ the inverted fermion matrix. The function  $C_B(x_2)$
is the correlator
$\lag\chb\chi(0)\chb\chi(x)\rag$ at zero spatial
momentum. The operator $\chb\chi(x)$ is local
and therefore very easy to measure, but has small overlap $Z_B$ with the bound
state we are actually looking for. One can write:
\beq
C_B(t)=C_{FF}(t) + Z_B\left(e^{-am_Bt}+e^{-am_B(T-t)}\right)
+ \tilde{Z}_B(-1)^t\left(e^{-a\tilde{m}_Bt}+e^{-a\tilde{m}_B(T-t)}\right)
~.
\lb{PROP:EQ}
\eeq
$C_{FF}$ is the free fermion--antifermion cut contribution
\be
C_{FF}(t)=\frac{1}{L}\sum_{x_1}G^{(2)}_F(0,x)G^{(2)}_F(0,x).
\lb{FER:CUT}
\ee
$G^{(2)}_F(0,x)$ is the free fermion propagator, the corresponding
$m_F$ and $Z_F$  are determined by fitting the measured fermion
propagator $\lag\chb(0)\chi(x)\rag$.

\begin{figure}
\centerline{
 \def\fpsangle{90}
 \epsfxsize=10cm
 \fpsbox{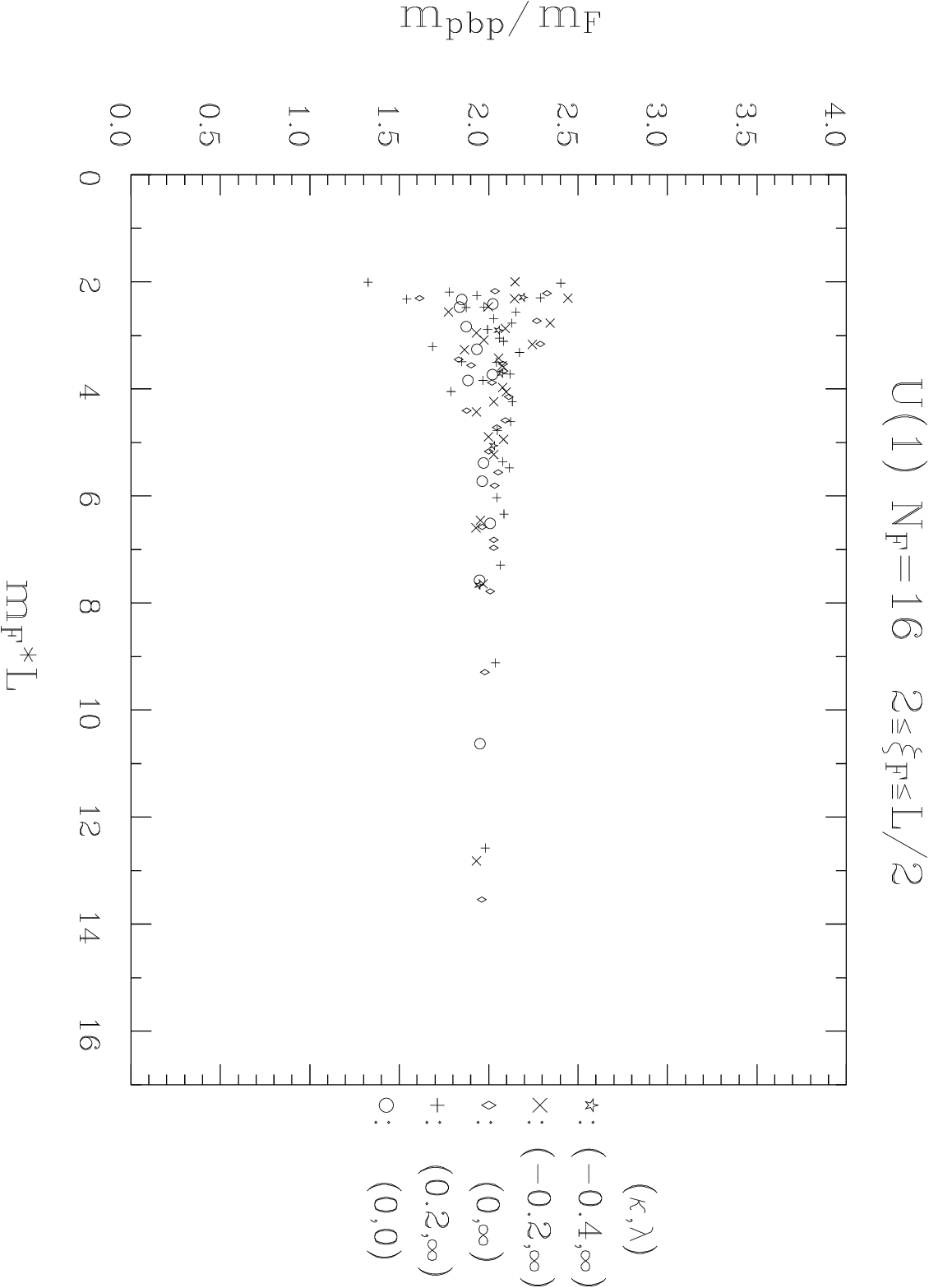}
}
\caption{The ratio $m_B/m_F$ plotted vs. $am_FL$. The results agree
  for all investigated $\kp$,$\lm$ pairs as long as the physical size
  of the lattice is sufficiently large, $am_FL>4$.
}
\label{RAT:FIG}
\end{figure}

The mass $am_B$ has been obtained at nearly all simulation points. The
results for those points, at which
\begin{equation}
2\leq\frac{1}{am_F}\leq\frac{L}{2}
\lb{RAT:SELECTION}
\end{equation}
are plotted in fig. \ref{RAT:FIG}. The first inequality in
(\ref{RAT:SELECTION}) selects points without significant lattice
artifacts, whereas the second one cuts off points  with too large
finite size effects. In fig. \ref{RAT:FIG} we display two physical,
dimensionless quantities: the ratio $m_B/m_F$ and $am_FL$, the size of
the system in the units of the fermionic correlation length.

The finite size effects increase as $am_FL$ decreases, and these
effects are not expected to be universal. But we see in fig.
\ref{RAT:FIG} that at least for $am_FL>4$, all the data are close to
the expected value $m_B/m_F=1.99$. This agreement
favours the conclusion that
for all shown ($\kp$,$\lm$) pairs, including
the GN limit, the theory is in the same universality class.

%
%

 The large uncertainties introduced by the inexact measurement of
$am_B$ do not allow to draw a definite conclusion, however.
 The cut contribution  $C_{FF}(t)$ in eq. (\eq{PROP:EQ}) is large:
about 95\%. This makes the determination of $m_B$ not very reliable.
We checked, however, that the fits for  $C_{B}(t)$ with only the cut
contribution are significantly worse than those with the full 
expression (\eq{PROP:EQ}). Nevertheless, we cannot exclude the
possibility that the result for the mass $m_B \simeq 2m_F$ 
 is an
artefact of the contribution of the fermion-antifermion threshold to
$C_{B}(t)$. In any case our results show that the mass of the lightest
fermion-antifermion bound state is not significantly lower than $2m_F$,
otherwise the mass $am_B$ would be easily measurable.
This still supports to some extent the universality
hypothesis.

\section{Transition between the Gross--Neveu and spin model universality
classes}

We have  tentatively concluded that the Y$_2$ models are equivalent to
the GN$_2$ models in the continuum limit $y\ra 0$ when $\kp$ is fixed at
some $\kp < \kp_c(\lm)$. This still allows that something new and
interesting happens in the limit when $y\ra 0$ and $\kp\ra\kp_c(\lm)$
simultaneously. If so, this would justify an independent existence of
the Y$_2$ models as field theories with physical content different from
the GN$_2$ universality class.

%
%
\begin{figure}
\centerline{
 \epsfxsize=7cm
 \epsfbox{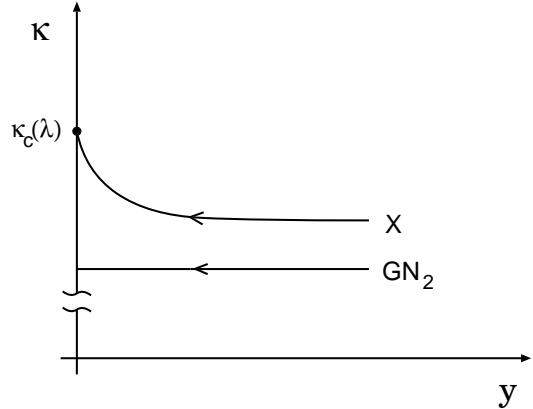}
}
 \caption{Two types of continuum limits from the $y>0$ in the vicinity
   of the line $\kp=\kp_c(\lm)$.
 }
\label{F:SKETCH}
\end{figure}

A hint that this could be the case is provided by the boson
spectrum considerations in the vicinity of the line $\kp=\kp_c(\lm)$ at
$y=0$. In fig. \ref{F:SKETCH} we indicate the situation at some fixed
$\lm$. The surface $y=0$, $\kp\leq\kp_c(\lm)$ corresponds to the high
temperature phase of
the scalar field theory which describes massive scalar particles 
of mass $m_\phi$.  
The continuum limit of this theory is obtained as
$\kp\ra\kp_c(\lm)_{-}$, where $am_\phi\ra 0$. The universality class for
$\lm>0$ is that of the 2D spin models with the same global symmetry as
the scalar field theory.
 The points $\kp<\kp_c(\lm)$ are regular points of the scalar field
theory, i.e. $am_\phi>0$. 
 Thus the end point of the line {\em GN$_2$} is a regular point and when
the continuum limit $am_F\ra 0$ is taken along this line, the ratio
$m_\phi/m_F$ becomes infinite. The scalar theory ``decouples'' and
the continuum GN$_2$ model results.

 The situation is different along certain lines of the type {\em X}
shown in fig. \ref{F:SKETCH}.
One can imagine a fine-tuned approach towards the line $\kp=\kp_c(\lm)$
at $y=0$ along which
$$
am_\phi,am_F \ra 0, ~~\frac{m_\phi}{m_F}\ra\alpha~,
$$
 with some arbitrary constant $\alpha$.  Assuming
applicability of the MF approximation, which can hold at least for
large $\al$ and not very close to the point $\kp=\kp_c$, one can even
calculate the corresponding paths (see Appendix).

 Thus the existence of a family of continuum field theories is
indicated, labeled by the ratio $\alpha=m_\phi/m_F$.
 Its spectrum might consist roughly of the same states as the GN$_2$
spectrum plus the spectrum of the scalar field theory in the high
temperature phase. We see no reason why these two ``sectors'' would not
interact with each other, so that these field theories might be of a new
kind, unexpected when the original action (\eq{CONTACTION}) of the Y$_2$
models is considered perturbatively. The situation is similar to
that found in \cc{CaEd94,So94} and we suggest that also in the case of
the Y$_2$ models there might be ``new universality classes which cannot
be seen in conventional perturbation theory''.

The new universality classes arising along the lines {\em X} could
perhaps be also  described by
other actions, different from 
(\eq{CONTACTION}), such that their physical content is more transparent.
A possible candidate would be the GN$_2$ model and an additional scalar
field theory in the high temperature phase, with both sectors coupled
to each other in such  a way that the spectrum of each sector does not
change much.
 As an example one could imagine the coupling of the form: 
$
             \psb F(\vr)\ps 
$
with $F(\vr)$ vanishing faster than $\vr$.

Unfortunately, we see little chance to investigate these issues
numerically. They constitute a part of the cross-over between two
distinctly different universality classes. But the cross-over phenomena
are notoriously difficult on finite lattices, as one has to work very
close to the critical point in order to disentangle various mixed
scaling tendencies. The difficulties we have encountered at $\lm=\infty$
and $\kp=0.2$ suggest that there we are already too close to the
crossover to be able to apply the simple MF method at those
modest correlation lengths we could afford.

\section{Discussion and conclusion}

We presented some numerical evidence that the Y$_2$ models belong to
the same universality class as the GN$_2$ models with the same chiral
symmetry and $N_F$ even if a strong scalar self-interaction prohibits
the use of the $1/N_F$ expansion. Of course, the universality, or
equivalence, of field theoretical models requires the equality of {\em
  all} observables like critical exponents, mass ratios, etc., in the
continuum limit. Thus our evidence, based mainly on the scaling
behaviour of the fermion mass, might seem to be only fragmentary.
However, the significance by our results is amplified by the 
agreement with the MF method. This method makes predictions of the
same type as the $1/N_F$ expansion in the lowest order, but has been
found to be applicable in a much wider range of parameters. The MF
method thus seems to be the suitable analytic method for studying the
Y$_2$ models, provided the pure scalar sector 
is taken into account fully, which is not difficult numerically
\cc{DeFo93b,St94}. The universality is a straightforward prediction of
this approach. 

 We found that the scaling law (\eq{Y2:SCALING}) also holds when the
bare kinetic term of the scalar field is negative. For $\kp < 0$ the
MF approximation works even better than for $\kp > 0$.
 We think that this
is due to a larger distance from the cross-over between the GN$_2$ and
the spin model universality classes. 
As the bare scalar field in 2D is dimensionless, the Y$_2$ models can be
chosen with quite arbitrary scalar field self-couplings and the
$\vr^4$ term, used in eq. (\eq{CONTACTION}), is only an example.
 However, we believe that
our investigation of the Y$_2$ models (\eq{ACTION}) with $\lm=\infty$,
i.e. $|\phi| = 1$ is generic 
for a broad class of the Y$_2$ models with quite arbitrary scalar
self-couplings. 

All that leads us to the conclusion that the Y$_2$ models belong to the
same universality class as the GN$_2$ models with the same chiral
symmetry and $N_F$ if $\kp < \kp_c$. Here $\kp_c$ is the boundary of
the high temperature phase of the pure scalar model with some rather
general self-couplings. The GN$_2$ models seem to be the most economic
representatives of these universality classes, as they do not contain
any ballast of irrelevant parameters. 

However, this conclusion does not imply that the Y$_2$ models are
completely superfluous. It could be that the continuum limits taken at
$\kp = \kp_c$ could lead to new universality classes, as some our
speculations indicate.

\appendix{}
\section{Appendix}

As mentioned in section 6 one can imagine that the ratio
$\alpha=m_\phi/m_F$ can be tuned when one approaches the
critical line $\kp=\kp_c(\lm)$, $y=0$.
 The value of $\alpha$ depends on the trajectory in
the bare parameter space along which the continuum limit is
approached.
 Let us consider a trajectory of the type X in figure \ref{F:SKETCH},
where $\kp_c(\lm)$ is approached from below.

 In the U(1) case at $\lm=\infty$ the limit $y=0$ is the well-known XY
model, with the scalar mass vanishing near $\kp=\kp_c$ and the
magnetic susceptibility diverging as:
\be
\chi \sim 1/ m_\phi^{2-\eta} \sim
\exp\left[b(\kp_c-\kp)^{-\nu}\right],~~~\nu=\frac{1}{2},~~~\eta=\frac{1}{4}.
\lb{CHI:APP}
\ee

The approach to $\kp=\kp_c(\lm)$, $\lm=\infty$ at fixed ratio
$\al=m_\phi/m_F$ is described by a function $y(\kp)$.  According to
the MF prediction (\eq{Y2:SCALING}) the fermion mass is given by:
\be
am_F \approx \mu_F\exp\left(-\frac{1}{2\bt_0}\frac{1}{\chi y^2}\right).
\lb{FER:MASS}
\ee
The ratio $\al$ now reads:
\be
\al \approx
\frac{m_\phi}{\mu_F}\exp\left( \frac{1}{2\bt_0}\frac{1}{ \chi y^2} \right), 
\ee
This can be solved for $y$:
\be
y^2(\kp) \approx \frac{1}{2\bt_0(\log({\al\mu_F})-\log{m_\phi})}
\cdot \frac{1}{\chi}
,
\lb{APP:SOL}
\ee
where the dependence of $\chi$ and $m_\phi$ on $\kappa$ is given by (\eq{CHI:APP}).
This formula can be used only as a rough approximation for the
trajectory. On can expect it to be applicable at most for
large $\alpha$ and not very close to $\kp=\kp_c$ (but close enough to
apply (\eq{CHI:APP})). 
Under these conditions eq. (\eq{APP:SOL}) can be simplified by keeping
only the leading exponential dependence
on $(\kp_c - \kp)$ in the r.h.s.:
\be
y^2(\kp)
\sim
\frac{1}{\log\al}\exp\left[-b(\kp_c-\kp)^{-\nu}\right]
.
\ee




\end{document}